\documentclass[paper]{ieice}
\usepackage{graphicx,xcolor}
\usepackage[fleqn]{amsmath}
\usepackage{newtxtext}
\usepackage[varg]{newtxmath}
\usepackage{subfigure}
\usepackage{url}
\usepackage[hang,small,bf]{caption}

\newcommand{\argmax}{\mathop{\rm argmax}\limits}

\field{}
\title{An Image Fusion Scheme for Single-Shot High Dynamic Range Imaging with Spatially Varying Exposures}
\authorlist{%
 \authorentry{Chihiro Go}{n}{affiliate label}\MembershipNumber{}
 \authorentry{Yuma Kinoshita}{s}{affiliate label}\MembershipNumber{1614118}
 \authorentry{Sayaka Shiota}{m}{affiliate label}\MembershipNumber{1410504}
 \authorentry[kiya@tmu.ac.jp]{Hitoshi Kiya}{f}{affiliate label}\MembershipNumber{7904021}
}
\affiliate[affiliate label]{The authors are with Department of Computer Science, Tokyo Metropolitan University, Tokyo, 191-0065 Japan.
}

\received{2019}{1}{1}
\revised{2019}{1}{1}

\begin{document}
\maketitle
\begin{summary}
This paper proposes a novel multi-exposure image fusion (MEF) scheme for single-shot high dynamic range imaging with spatially varying exposures (SVE).
Single-shot imaging with SVE enables us not only to produce images without color saturation regions from a single-shot image, but also to avoid ghost artifacts in the producing ones.
However, the number of exposures is generally limited to two, and moreover it is difficult to decide the optimum exposure values before the photographing.
In the proposed scheme, a scene segmentation method is applied to input multi-exposure images, and then the luminance of the input images is adjusted according to both of the number of scenes and the relationship between exposure values and pixel values.
The proposed method with the luminance adjustment allows us to improve the above two issues.
In this paper, we focus on dual-ISO imaging as one of single-shot imaging.
In an experiment, the proposed scheme is demonstrated to be effective for single-shot high dynamic range imaging with SVE, compared with conventional MEF schemes with exposure compensation.

\end{summary}

\section{Introduction}
The low dynamic range (LDR) imaging sensors used in modern digital cameras cannot capture the wide dynamic range of a real scene.
The limitation causes blocked up shadows and blown out highlights in images taken by digital cameras.
In addition, those images often have low contrast.
For this reason, a lot of high dynamic range (HDR) imaging techniques have so far been reported.

The most common approach for HDR imaging is multi-exposure image fusion (MEF).
By using the approach, images covering the HDR of real scenes are generated
by fusing a set of differently exposed images called as multi-exposure images.
To produce high-quality fused images, three or more multi-exposure images are generally utilized as inputs for multi-exposure fusion.
However, MEF often causes ghost like artifacts in fused images.
This is because the movement of cameras and subjects makes it difficult to capture suitable multi-exposure images.
While there are various robust fusion methods against ghost like artifacts \cite{ghost1,ghost2,kinoshita2017pseudo,kinoshita2018pseudo}, the performance is still limited.

For capturing images covering the HDR without the issue of ghost like artifacts, camera devices having a wide dynamic image sensor \cite{tocci2011versatile} or spatially varying exposures (SVE) have been studied.
Although the former devices are very expensive and are not widespread yet, the latter ones can be applied to commonly used digital cameras.
In the SVE-based imaging, an image is captured with a single shutter by varying exposures for each pixel on an imaging sensor, and multiple sub-images are obtained by separating the image for each exposure.
Varying exposures is done by spatially changing shutter speeds or ISO speeds.
The dual-ISO imaging is one of the SVE-based imaging methods, in which varying the ISO speeds alternates every two lines in a single raw Bayer image \cite{alex,hajsharif2014hdr,gil2016high}.
In \cite{sve1,intro2,AnL17}, the exposure time alternates row-wise varying exposures in a single raw Bayer image with two exposure times.
In Quad Bayer pixel structure, integration can be divided into long-time integration and short-time integration for every two pixels in the Quad array \cite{quadbayer}.
However, in these methods, there is a trade-off between the number of exposures and the resolution of captured sub-images.
Hence, the number of sub-images, which is generally two, is less than that of images used for conventional MEF methods.
For this reason, the scene information cannot be sufficiently expressed in resulting images.

Because of such a situation, we propose a new image fusion method for single-shot imaging with SVE.
The proposed method generates more multi-exposure images from two images captured by SVE, and fuse them into a single high-quality image.
To generate those images, a new scene segmentation method is applied to input multi-exposure images.
After that, the exposure compensation for input images is automatically performed so that generated multi-exposure images clearly show all regions in a scene \cite{kinoshita2018automatic,kinoshita2018multi,kinoshita_kiya_2018,kinoshita2019scene}.
Generated multi-exposure images can be applied to any MEF methods.

We evaluate the effectiveness of the proposed method in terms of the quality of generated images by two simulations.
In the simulations, the proposed method is compared with conventional MEF methods in terms of objective quality metrics: the tone mapped image quality index (TMQI) \cite{yeganeh2013objective}, MEF structural similarity (MEF-SSIM) \cite{ma2015perceptual}, statistical naturalness, and discrete entropy.
Experimental results show that the proposed method can produce high-quality images compared with conventional fusion methods for single-shot high dynamic range imaging with SVE.

\section{Preparation}
\label{sec:preparation}
Here we summarize dual-ISO imaging and a conventional fusion scheme for the imaging.
\renewcommand{\thesubsubsection}{\Alph{subsubsection}}
\subsection{Dual-ISO imaging}
Sony Corp. provides an imaging sensor product which can take SVE images with the Quad Bayer array \cite{quadbayer}.
Canon Inc. also provides some cameras which can capture SVE images by changing the ISO speed of the image sensor line by line by using firmware, called Magic Lantern \cite{alex}.
SVE images are generally expressed as a row image format.
Here, we focus on dual-ISO imaging as Canon's one.

\begin{figure}[t]
	\begin{center}
	\includegraphics[width=50mm]{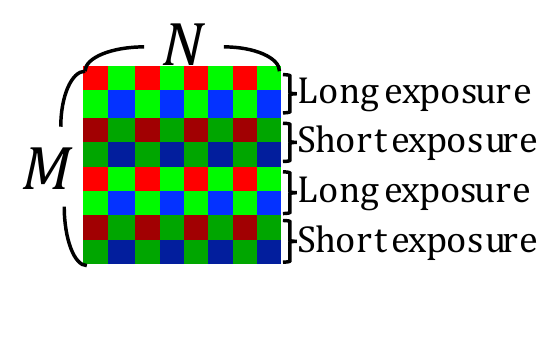}
	\end{center}
	\caption{Raw Bayer image sensed with dual-ISO sensor}
\label{fig:du_sen}
\end{figure}

\begin{figure*}[t]
	\begin{center}
	\includegraphics[width=140mm]{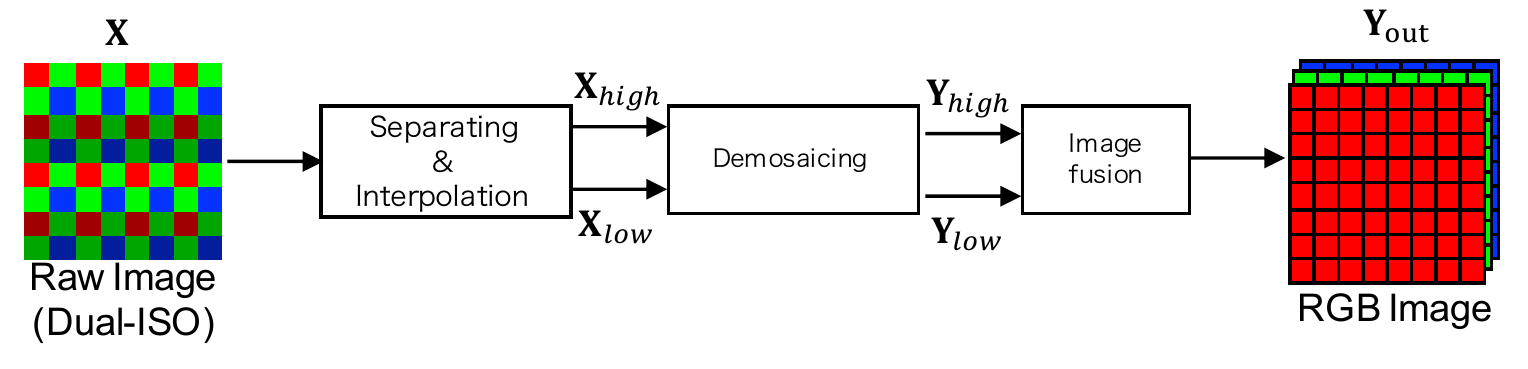}
	\end{center}
	\caption{Conventional method for dual-ISO imaging}
\label{fig:al}
\end{figure*}

A raw Bayer image sensed with a dual-ISO sensor is illustrated in Fig.\ref{fig:du_sen}, where the ISO speed alternates every two lines in the Bayer image \cite{alex}.
By using the dual-ISO sensor, raw images with two exposures are produced.
The Bayer image captured with two ISO speeds are fused as shown in Fig.\ref{fig:al}.
Each fusion step in Fig.\ref{fig:al} is briefly explained as below.

\subsubsection{Separation and interpolation}
\begin{figure}[t]
	\centering
	\subfigure[Separation]{
		\label{subfig:separation}
		\includegraphics[width = 0.45\columnwidth]{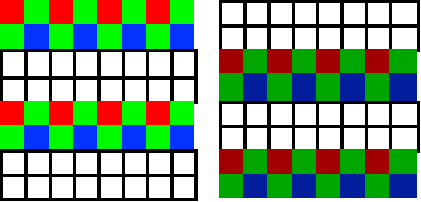}}
	\subfigure[Interpolation]{
		\label{subfig:interpolation}
		\includegraphics[width = 0.45\columnwidth]{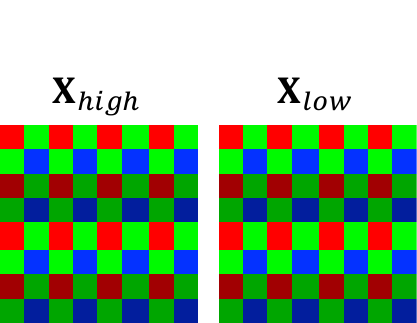}}
	\caption{Separating and Interpolation}\label{fig:sepint}
\end{figure}

A raw image $\bf X$ with a size of $M\times N$ is first divided into two raw images with the size of $M/2 \times N$, according to the difference of ISO speed (See Fig.\ref{fig:sepint}).

Next, interpolation processing is applied to each raw image for producing two raw images with the size of $M \times N$ : ${\bf X}_{low}$ and ${\bf X}_{high}$.

\subsubsection{Demosaicing}
To obtain two RGB images ${\bf Y}_{low}$, ${\bf Y}_{high}$, an image demosaicing algorithm is applied to two raw images ${\bf X}_{low}, {\bf X}_{high}$.

\subsubsection{Image fusion}
A fused image ${\bf Y}_{\rm out}$ is produced as
\begin{equation}
	{\bf Y}_{\rm out} = \mathcal{F}({\bf Y}_{low},{\bf Y}_{high}),
	\label{eq:image_fusion}
\end{equation}
where $\mathcal{F}(\cdot)$ indicates a fusion to fuse two images into a single image.

In conventional single-shot imaging, the number of exposures is generally limited to two, and moreover it is difficult to decide the optimum exposure values before the photographing as described above.
We aim to improve the above two issues.

\section{Proposed method}
\label{sec:proposed}
\begin{figure*}[t]
	\begin{center}
	\includegraphics[width = 1.8\columnwidth]{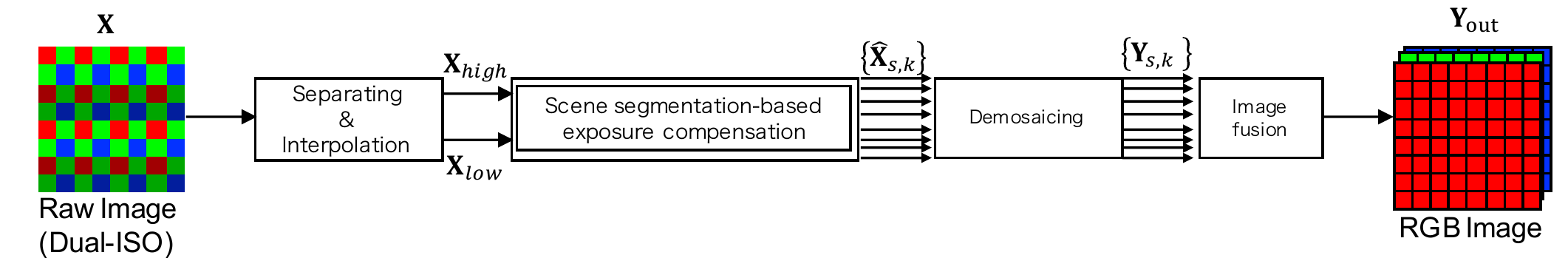}
	\end{center}
	\caption{Outline of proposed method}
\label{fig:proposed_flow}
\end{figure*}

\begin{figure*}[t]
	\begin{center}
	\includegraphics[width = 1.8\columnwidth]{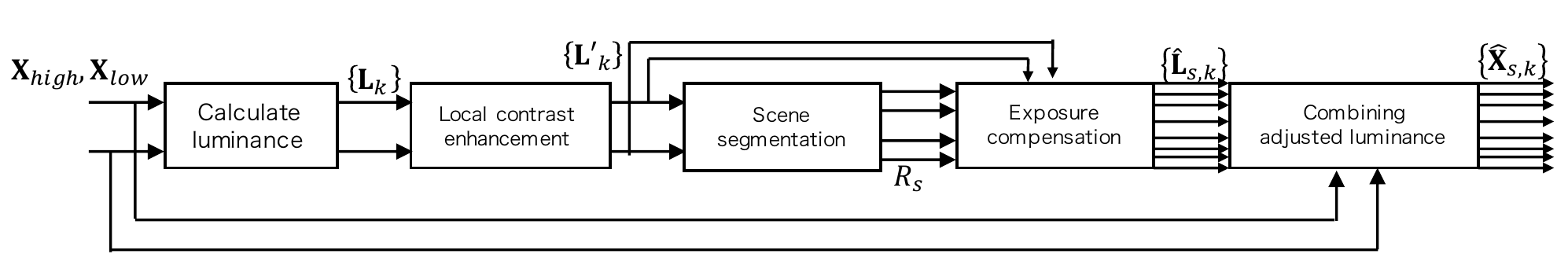}
	\end{center}
	\caption{Details of scene segmentation-based exposure compensation}
\label{fig:proposed_detail}
\end{figure*}

In order to improve the two issues that the conventional single-shot imaging has, we propose a novel image fusion method for the single-shot imaging.
The outline of the proposed method is shown in Fig.\ref{fig:proposed_flow}, where the main contribution of this work is in scene-segmentation based exposure competition.
The exposure competition consists of the following five steps (See Fig.\ref{fig:proposed_detail}).

\subsubsection{Local contrast enhancement}
\begin{figure}[t]
	\centering
	\subfigure[Images without local contrast enhancement]{
		\label{subfig:le-bf}
		\includegraphics[width = \columnwidth]{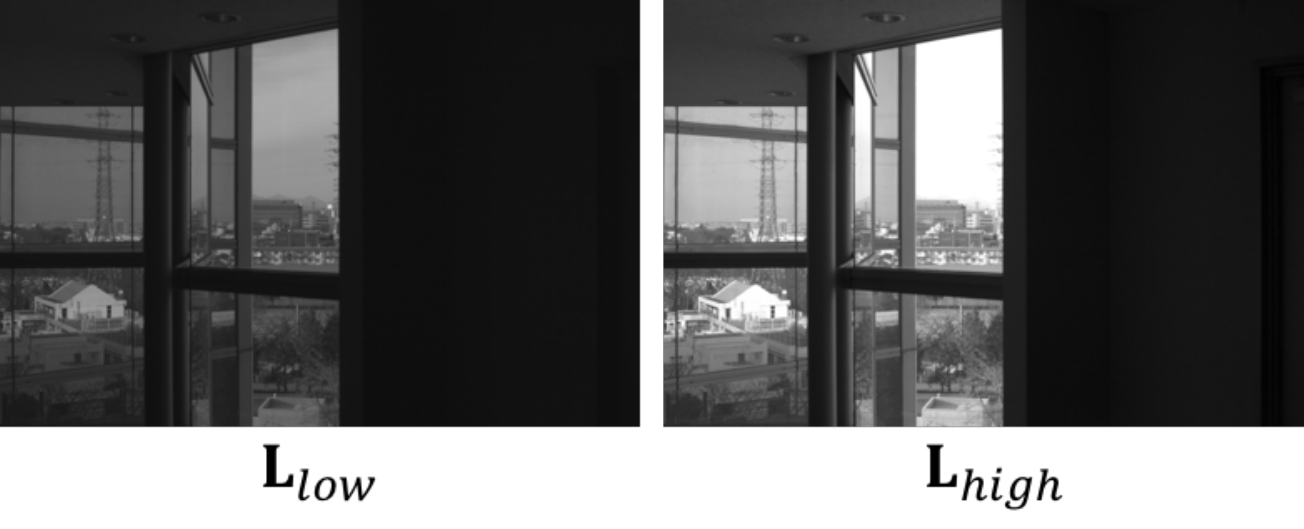}}
	\subfigure[Images with local contrast enhancement]{
		\label{subfig:le-af}
		\includegraphics[width = \columnwidth]{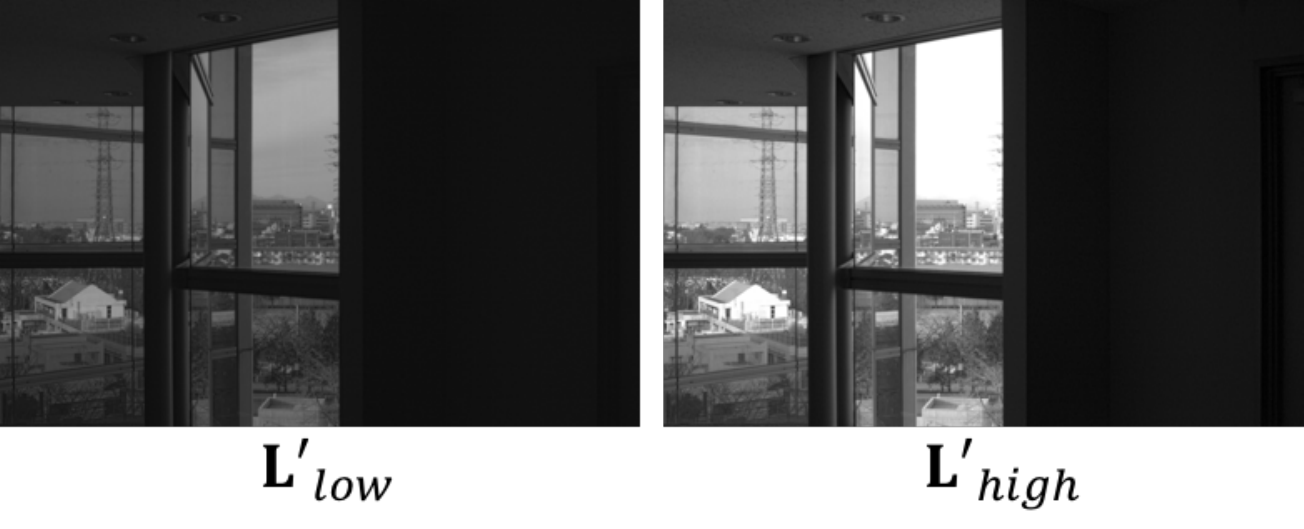}}
	\hspace{20mm}
	\subfigure[Scene segmentation result]{
		\label{subfig:hdr-b}
		\includegraphics[width = 0.6\columnwidth]{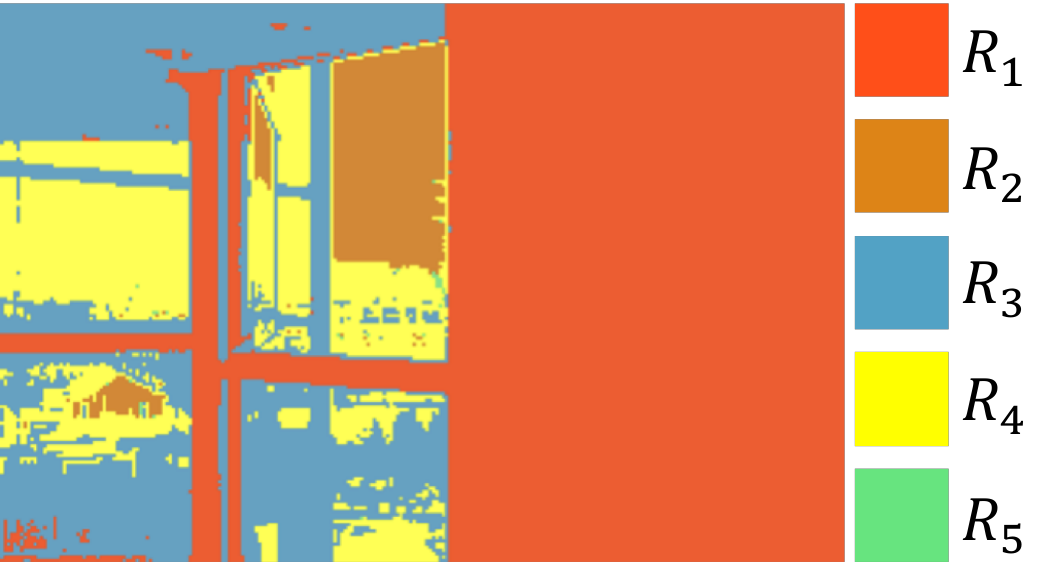}}
	\caption{Example of scene segmentation}
	\label{fig:ss_ex}
\end{figure}

Since the number of exposures is generally limited to two, ${\bf X}$ cannot always represent a scene clearly, unlike general multi-exposure images.
A local contrast enhancement algorithm is used to enhance detailed information in ${\bf X}$.
In this paper, the local contrast enhancement using the dodging and burning algorithm \cite{huo2013dodging} is performed as
\begin{equation}
	L_k'(i,j)=\frac{L_k^2(i,j)}{L_{ak}(i,j)},\:k\in \{low,~high\},
	\label{eq:local_enhance}
\end{equation}
where $L_k(i,j)$ is the luminance value of ${\bf X}_k,k\in\{low,high\},$ at the place $(i,j)$, and $L_{ak}(i,j)$ is the local average of luminance $L_k(i,j)$ around pixel $(i,j)$.
Here, a bilateral filter is performed to obtain $L_{ak}(i,j)$ as in \cite{huo2013dodging}.
In Fig.\ref{subfig:le-bf}, images with the local enhancement are compared with images without any local enhancement.
In contrast, Fig.\ref{subfig:le-af} shows images with the local enhancement.

\begin{figure}[t]
	\centering
	\subfigure[Red]{
		\label{subfig:raw2lum_r}
		\includegraphics[width = 0.45\columnwidth]{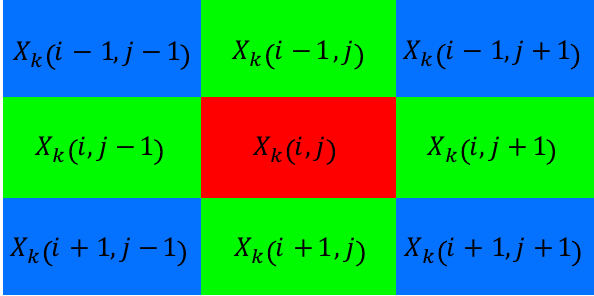}}
	\subfigure[Green1]{
		\label{subfig:raw2lum_g1}
		\includegraphics[width = 0.45\columnwidth]{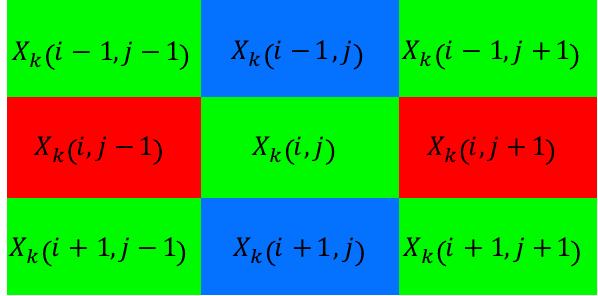}}
	\hspace{20mm}
	\subfigure[Green2]{
		\label{subfig:raw2lum_g2}
		\includegraphics[width = 0.45\columnwidth]{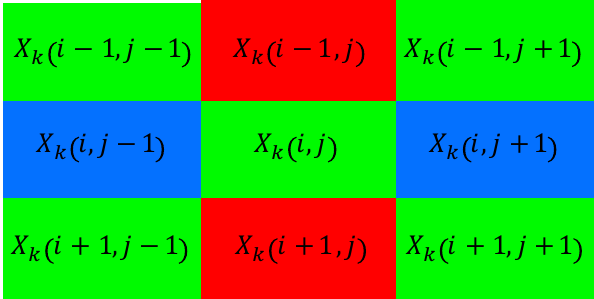}}
	\subfigure[Blue]{
		\label{subfig:raw2lum_b}
		\includegraphics[width = 0.45\columnwidth]{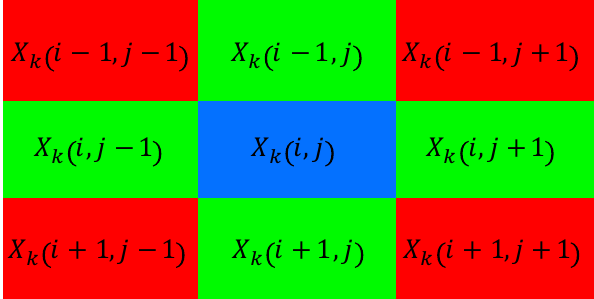}}
		\caption{Block for luminance calculation}
	\label{fig:raw2lum}
\end{figure}

The luminance ${\bf L}_k=\{L_k(i,j)\}$ at all pixels is required to calculate eq.\eqref{eq:local_enhance}, but a pixel value of a raw image ${\bf X}_k,X_k(i,j),$ has only a red, green or blue value.
In this paper, in order to obtain luminance ${\bf L}_k$ from ${\bf X}_k$, $L_k(i,j)$ is calculated by using $X_k(i,j)$ and its eight surrounding pixels as shown in Fig.\ref{fig:raw2lum}.
When $X_{k}(i,j)$ is a red value as in Fig.\ref{subfig:raw2lum_r}, the luminance $L_k(i,j)$ is calculated, as
\begin{eqnarray}
\scalebox{0.85}{$\displaystyle
L_{k}(i,j)=0.27X_{R,k}(i,j)+0.67X_{G,k}(i,j)+0.06X_{B,k}(i,j),
$}
\label{eqn:x2lum}
\end{eqnarray}
where 
\begin{equation}
	\label{eq:x2lum_R}
	\begin{split}
	X_{R,k}(i,j)&=X_k(i,j)\\
	X_{G,k}(i,j)&=\frac{1}{4}\sum_{x=\pm1} (X_k(i+x,j)+X_k(i,j+x))\\
	X_{B,k}(i,j)&=\frac{1}{4}\sum_{x=\pm1,y=\pm1} X_k(i+x,j+y).
	\end{split}
\end{equation}
When $X_{k}(i,j)$ is a blue value as in Fig.\ref{subfig:raw2lum_b}, the luminance $L_k(i,j)$ is calculated, as
\begin{equation}
	\label{eq:x2lum_B}
	\begin{split}
	X_{R,k}(i,j)&=\frac{1}{4}\sum_{x=\pm1,y=\pm1} X_k(i+x,j+y)\\
	X_{G,k}(i,j)&=\frac{1}{4}\sum_{x=\pm1} (X_k(i+x,j)+X_k(i,j+x))\\
	X_{B,k}(i,j)&=X_k(i,j).
	\end{split}
\end{equation}
Similarly, when $X_{k}(i,j)$ is a green value as in Fig.\ref{subfig:raw2lum_g1} or \ref{subfig:raw2lum_g2}, the luminance $L_k(i,j)$ is calculated the same way, respective.
Eqs. \eqref{eq:x2lum_R} and \eqref{eq:x2lum_B} are a simple demosaicing algorithm.
Other demosaicing ones can be applied to ${\bf X}_{low}$ and ${\bf X}_{high}$.

\subsubsection{Scene segmentation}
The goal of the proposed segmentation is to separate images into $S$ areas ${\rm R}_1,\ldots,{\rm R}_S\subset {\rm R}$, where each of them has a specific brightness range of the image and satisfies ${\rm R}_1 \cup {\rm R}_2 \cup \cdots \cup {\rm R}_S = {\rm R}$.
These results are used for exposure compensation.

The proposed segmentation method differs from typical segmentation ones in two ways.
\begin{itemize}
  \item Drawing no attention to the structure of images, e.g., edges.
  \item Allowing ${\rm R}_s$ to include spatially non-contiguous regions.
\end{itemize}
For the segmentation, a Gaussian mixture distribution is utilized to model the luminance distribution of the input images in this paper. 
After that, pixels are classified by using a clustering algorithm based on a Gaussian mixture model (GMM).

To obtain a model considering the luminance values ${\bf L}'_{high}$ and ${\bf L}'_{low}$, we regard luminance values at a pixel $(i,j)$ as a 2-dimensional vector ${\bf l}(i,j) = \{L'_{low}(i,j), L'_{high}(i,j)\}^{\mathrm{T}}$, where $T$ denotes the transpose of a vector.
By using a GMM, the distribution of ${\bf l}(i,j)$ is given as
\begin{equation}
	p\left({\bf l}(i,j)\right) = \sum_{d=1}^D\pi_dG\left({\bf l}(i,j)|{\boldsymbol \mu}_d,{\bf \Sigma}_d\right),
\end{equation}
where $D$ indicates the number of mixture components, $\pi_d$ is the $d$th mixing coefficient, and $G\left({\bf l}(i,j)|{\boldsymbol \mu}_d,{\bf \Sigma}_d\right)$ is a 2-dimensional Gaussian distribution with mean ${\boldsymbol \mu}_d$ and variance covariance matrix ${\bf \Sigma}_d$.

To fit the GMM into given ${\bf l}(i,j)$, the variational Bayesian algorithm \cite{Bishop:2006:PRM:1162264} is utilized.
Compared with the traditional maximum likelihood approach, one of the advantages is that the variational Bayesian approach can avoid overfitting even when we choose a large $D$.
For this reason, unnecessary mixture components are automatically removed by using the approach together with a large $D$.
$D=10$ is used in this paper, as the maximum of the partition number $S$.

Here, let $z$ be a $D$-dimensional binary random variable having a 1-of-$D$ representation in which a particular element $z_d$ is equal to 1 and all other elements are equal to 0. The marginal distribution over $z$ is specified in terms of a mixing coefficient $\pi_d$, such that
\begin{equation}
	p(z_d = 1) = \pi_d.
\end{equation}
For $p(z_d=1)$ to be a valid probability, $\{\pi_d\}$ must satisfy
\begin{equation}
	0\leq \pi_d \leq 1
\end{equation}
together with
\begin{equation}
	\sum_{d=1}^D\pi_d = 1.
\end{equation}

A cluster for an observation ${\bf l}(i,j)$ is determined by the responsibility $\gamma\left(z_d|{\bf l}(i,j)\right)$ which is given as a conditional probability:
\begin{equation}
	\begin{split}
	\gamma\left(z_d|{\bf l}(i,j)\right)&=p(z_d=1|{\bf l}(i,j))\\
	&= \frac{\pi_d G({\bf l}(i,j)|{\boldsymbol \mu}_d,\Sigma_d)}{\sum_{j=1}^D\pi_j G({\bf l}(i,j)|{\boldsymbol \mu}_j,\Sigma_j)}.
	\end{split}
\end{equation}
When a pixel $(i,j)\in {\rm R}$ is given and $s$ satisfies
\begin{equation}
	s = \argmax_d\gamma(z_d|{\bf l}(i,j)),
\end{equation}
the pixel $(i,j)$ is assigned to a subset ${\rm R}_s$ of ${\rm R}$.
Figure \ref{fig:ss_ex} shows an example of scene segmentation.
In this example, the image was segmented into $S=5$ parts.

\subsubsection{Exposure compensation}
$2S$ multi-exposure images are created from two images, ${\bf L}'_{low}, {\bf L}'_{high}$, by using the result of scene segmentation, .
The scaled luminance ${\bf\hat L}_{s,k}$ which clearly represents an area ${\rm R}_s$ is obtained by
\begin{equation}
	{\hat L}_{s,k}(i,j) = \alpha_{s,k} L_k'(i,j),
	\label{eq:ec}
\end{equation}
where the scale factor $\alpha_{s,k}>0$ indicates the degree of adjustment for the $s$th scaled luminance ${\bf\hat L}_{s,k}$.
In the following, how to determine parameter $\alpha_{s,k}$ is discussed.

Given ${\rm R}_s$ as a subset of ${\rm R}$, the approximate brightness of an area ${\rm R}_s$ is calculated as the geometric mean of luminance values on ${\rm R}_s$.
We thus estimate an adjusted multi-exposure image ${\hat L}_{s,k}(i,j)$, so that the geometric mean of its luminance equals to middle-gray of the displayed image, or 0.18 on a scale from zero to one as in \cite{reinhard2002photographic}.

The geometric mean $g(L_k'|{\rm R}_s)$ of luminance $L_k'$ on pixel set ${\rm R}_s$ is calculated using
\begin{equation}
	g(L_k'|{\rm R}_s) = \exp\left(\frac{1}{|{\rm R}_s|}\sum_{(i,j)\in {\rm R}_s}\ln(\max(L_k'(i,j),\epsilon))\right),
	\label{eq:geometric}
\end{equation}
where $\epsilon$ is set to a small value to avoid singularities at $L_k'(i,j)=0$.
From eq.\eqref{eq:geometric}, parameter $\alpha_{s,k}$ is calculated as
\begin{equation}
	\alpha_{s,k} = \frac{0.18}{g(L_k'|{\rm R}_s)}.
	\label{eq:alpha}
\end{equation}
By using the exposure compensation, exposure values are automatically adjusted, even when the values have no approximate brightness.

\subsubsection{Combining adjusted luminance and input images}
A set $\{{\bf\hat L}_{s,k}\}$ of luminance adjusted by the scene segmentation-based exposure compensation is combined with an input image $\{{\bf X}_k\}$ to obtain adjusted images $\{{\bf \hat X}_{s,k}\}$.

Therefore, the adjusted pixel value ${\hat X}_{s,k}(i,j)$ is computed by
\begin{equation}
	\begin{split}
	&\hat X_{s,k}(i,j) = \frac{{\hat L}_{s,k}(i,j)}{L_k(i,j)}X_k(i,j),\\
	&(s = 1,2,\ldots,S, ~ k\in\{low,\:high\}).
	\end{split}
	\label{eq:adjusted image}
\end{equation}
As a result, $2S$ raw images $\{\hat{\bf X}_{s,k}\}$ are prepared, according to eq.\eqref{eq:adjusted image}.

\subsubsection{Demosaicing}
Since $\{{\bf \hat{X}}_{s,k}\}$ are Raw images, a demosaicing algorithm is carried out to obtain $2S$ RGB images $\{{\bf Y}_{s,k}\}$.
In this paper, we apply an image demosaicing algorithm \cite{demosaicing} to $\{{\bf \hat{X}}_{s,k}\}$.
In the proposed method, other demosaicing ones can be applied to not only ${\bf X}_{low}$ and ${\bf X}_{high}$ but also $\{{\bf \hat{X}}_{s,k}\}$.
Demosaicing algorithms give some influence to the quality of generated images such as the presence of artifacts.

\subsubsection{Image fusion}
A final image $\bf{Y}_{\rm out}$ is produced by using $2S$ RGB images, as
\begin{equation}
{\bf Y}_{\rm out} = \mathcal{F}({\bf Y}_{1,low},\ldots,{\bf Y}_{S,low},{\bf Y}_{1,high},\ldots,{\bf Y}_{S,high}),
\end{equation}
where $\mathcal F (\cdot)$ indicates a function to fuse multi-exposure images into a single image.
Any existing MEF methods are applicable for the proposed scheme.
The fusion method proposed by Mertens et al. \cite{mertens2009exposure} is used in this paper as $\mathcal F(\cdot)$.

\section{Simulation}
In experiments, the proposed scheme is demonstrated to be effective for single-shot HDR imaging with SVE.
The performance of the proposed scheme was compared with conventional MEF methods \cite{alex,yang2018multi}.

\subsection{Simulation with HDR images}
\subsubsection{Dataset}
In this experiment, input SVE images ${\bf X}$ were generated
by using HDR images ${\bf Y}_{\rm HDR}$,
according to \cite{reinhard2002photographic}.
The procedure for generating ${\bf X}$ consists of
three steps.
First, an LDR image ${\bf Y}_{\rm 0EV}$ with $0$ EV
was generated from an HDR image,
so that the geometric mean of its luminance
equals 0.18, as in \cite{reinhard2002photographic}.
Next, according to the following equation,
exposure values of every two lines in ${\bf Y}_{\rm 0EV}$
were changed into $k$ EV and $-k$ EV, respectively
\begin{equation}
\begin{split}
Y(i,j) =
\begin{cases}
2^kY_{\rm 0EV}(i,j) & i = 4m-3,4m-2 \\
2^{-k}Y_{\rm 0EV}(i,j) & otherwise
\end{cases}
,\\
m=1,2,\ldots,M/4,~ j=1,2,\ldots,N.
\end{split}
\end{equation}
Finally, raw Bayer image ${\bf X}$ was
obtained by removing two of RGB components at each pixel
in $Y(i,j)$.
To generate ${\bf X}$,
we used 28 HDR images which were selected from a database \cite{hdrlabs}.

From each HDR image, four SVE image sets with $\pm 1$ EV, $\pm 2$ EV, $\pm3$ EV, or $\pm4$ EV were generated.
In MEF, it is not known yet how the optimal EV is decided.
Therefore, in MEF, this issue is overcome by capturing many EV images.
However, for SVE images, many EV images can not be captured.
The proposed method aims to improve this issue under the limited number of exposure values.

\subsubsection{Objective metrics}
The quality of images produced by each method was evaluated in two objective metrics; TMQI \cite{yeganeh2013objective} and MEF-SSIM \cite{ma2015perceptual}.
TMQI measures the quality of a tone mapped image from an HDR image and it consists of structural fidelity and statistical naturalness.
To calculate structural fidelity, an HDR image is used as a reference.
In contrast, statistical naturalness does not need any references.
MEF-SSIM is based on a multi-scale SSIM framework and a patch consistency measure.
It keeps a good balance between local structure preservation and global luminance consistency.
For each score, a larger value means higher quality.

\subsubsection{Result}
Tables \ref{tb:tmqi_all} and \ref{tb:ssim_all} denote the average scores of 28 images.
From Table \ref{tb:tmqi_all}, it is confirmed that the proposed method had higher TMQI scores than conventional methods.
From Table.\ref{tb:ssim_all}, in the case of $\pm3$ EV and $\pm4$ EV, the proposed method had higher scores than the conventional methods, although Yang et al. had higher MEF-SSIM scores than the proposed method in the case of $\pm1$ EV and $\pm2$ EV.
In addition, the difference between with the local enhancement and without any local enhancement is demonstrated in the tables.
In particular, the image quality was improved by performing the local contrast enhancement in terms of MEF-SSIM.
In addition, we confirmed the influence of demosaicing algorithms.
For "proposed (with Eqs. \eqref{eq:x2lum_R} and \eqref{eq:x2lum_B})" in the tables, Eqs. \eqref{eq:x2lum_R} and \eqref{eq:x2lum_B} were applied to not only ${\bf X}_{low}$ and ${\bf X}_{high}$ but also $\{{\bf \hat{X}}_{s,k}\}$.
The results were lower than other proposed ones with the demosaicing algorithm \cite{demosaicing}, although the results were higher than those of conventional MEF methods.

\begin{figure*}[t]
	\centering
	\subfigure[No exposure compensation ($\pm 1\rm EV$)]{
		\label{subfig:hdr-b}
		\includegraphics[width = 0.5\columnwidth]{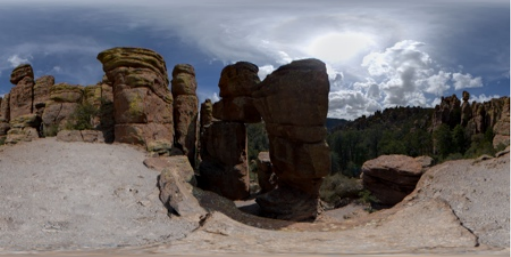}}
	\subfigure[Alex ($\pm 1\rm EV$)]{
		\label{subfig:hdr-c}
		\includegraphics[width = 0.5\columnwidth]{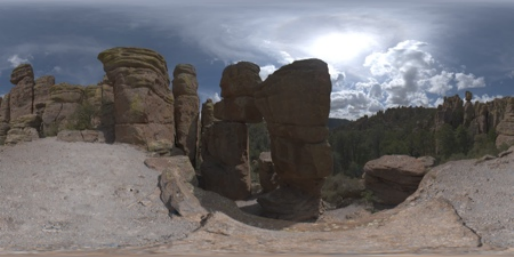}}
	\subfigure[Yang et al. ($\pm 1\rm EV$)]{
		\label{subfig:hdr-d}
		\includegraphics[width = 0.5\columnwidth]{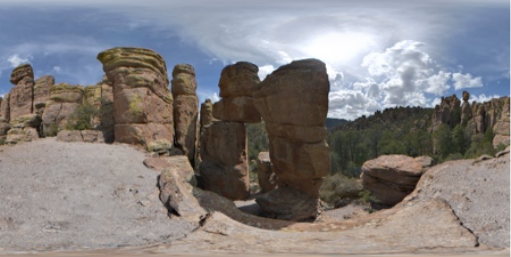}}
	\subfigure[Proposed method ($\pm 1\rm EV$)]{
		\label{subfig:hdr-a}
		\includegraphics[width = 0.5\columnwidth]{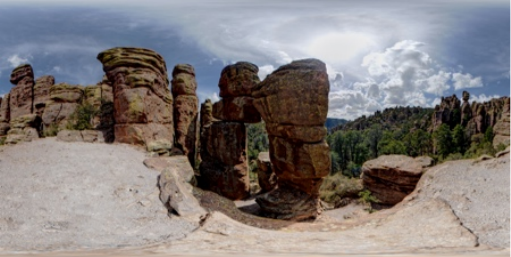}}
	\hspace{20mm}
	\subfigure[No exposure compensation ($\pm 2\rm EV$)]{
		\label{subfig:hdr-f}
		\includegraphics[width = 0.5\columnwidth]{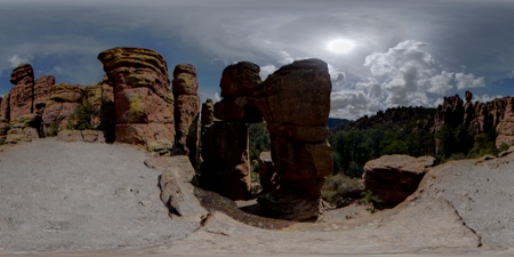}}
	\subfigure[Alex ($\pm 2\rm EV$)]{
		\label{subfig:hdr-g}
		\includegraphics[width = 0.5\columnwidth]{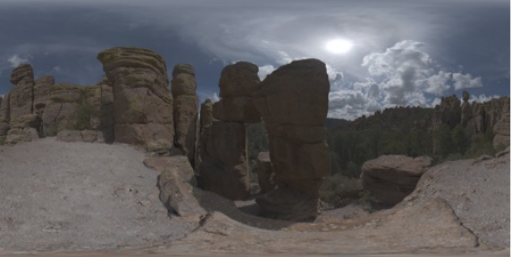}}
	\subfigure[Yang et al. ($\pm 2\rm EV$)]{
		\label{subfig:hdr-h}
		\includegraphics[width = 0.5\columnwidth]{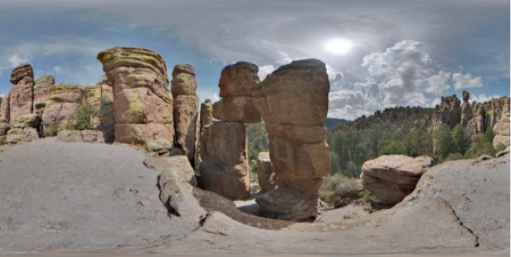}}
	\subfigure[Proposed method ($\pm 2\rm EV$)]{
		\label{subfig:hdr-e}
		\includegraphics[width = 0.5\columnwidth]{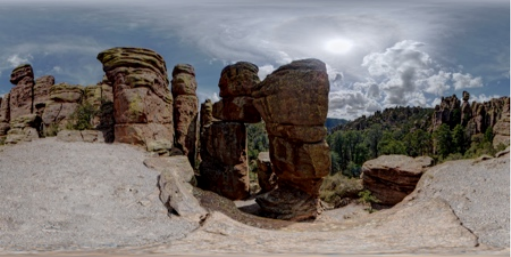}}
	\hspace{20mm}
	\subfigure[No exposure compensation ($\pm 3\rm EV$)]{
		\label{subfig:hdr-j}
		\includegraphics[width = 0.5\columnwidth]{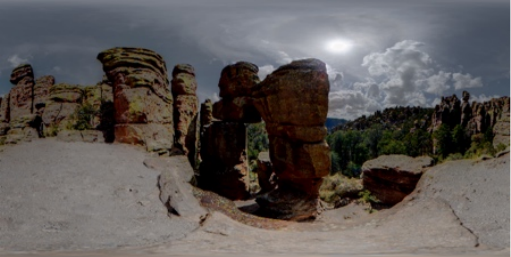}}
	\subfigure[Alex ($\pm 3\rm EV$)]{
		\label{subfig:hdr-k}
		\includegraphics[width = 0.5\columnwidth]{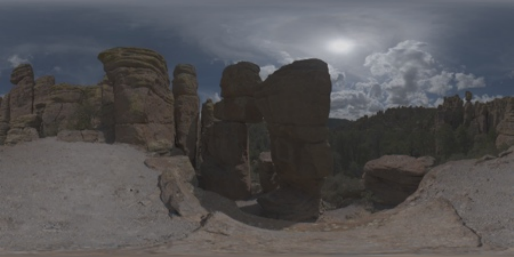}}
	\subfigure[Yang et al. ($\pm 3\rm EV$)]{
		\label{subfig:hdr-l}
		\includegraphics[width = 0.5\columnwidth]{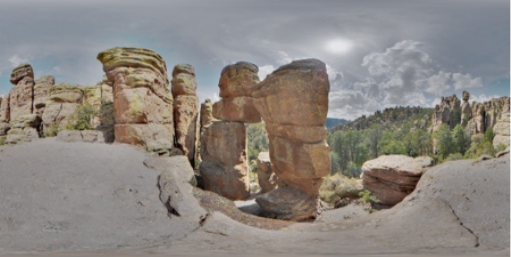}}
	\subfigure[Proposed method ($\pm 3\rm EV$)]{
		\label{subfig:hdr-i}
		\includegraphics[width = 0.5\columnwidth]{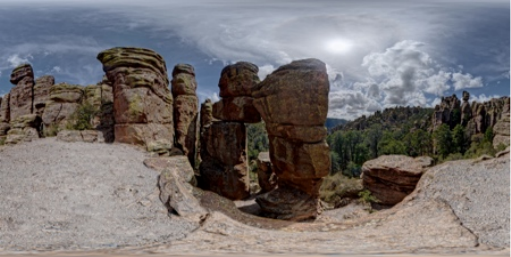}}
	\hspace{20mm}
	\subfigure[No exposure compensation ($\pm 4\rm EV$)]{
		\label{subfig:hdr-n}
		\includegraphics[width = 0.5\columnwidth]{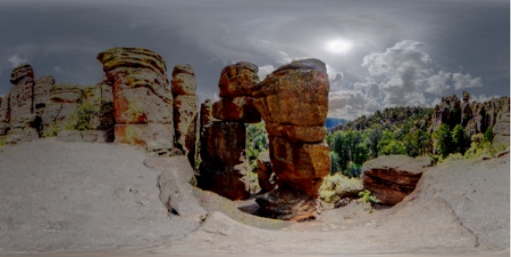}}
	\subfigure[Alex ($\pm 4\rm EV$)]{
		\label{subfig:hdr-o}
		\includegraphics[width = 0.5\columnwidth]{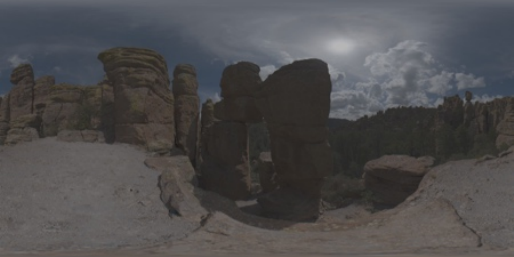}}
	\subfigure[Yang et al. ($\pm 4\rm EV$)]{
		\label{subfig:hdr-p}
		\includegraphics[width = 0.5\columnwidth]{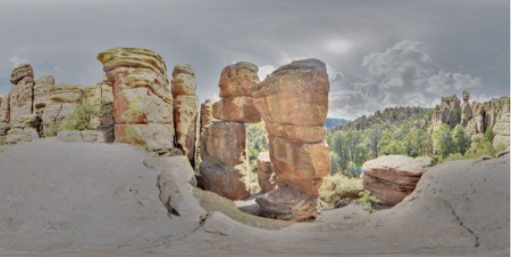}}
	\subfigure[Proposed method ($\pm 4\rm EV$)]{
		\label{subfig:hdr-m}
		\includegraphics[width = 0.5\columnwidth]{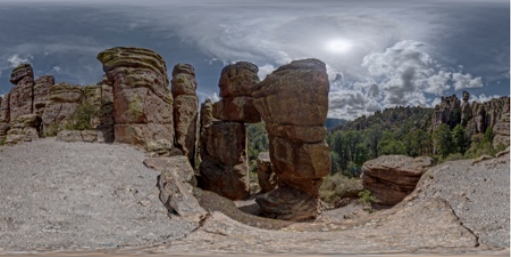}}
	\caption{Examples of fused images ${\bf Y}_{\rm out}$ $(S=6)$}
	\label{fig:result_1}
\end{figure*}

Figure \ref{fig:result_1} shows examples of output images ${\rm Y}_{out}$ produced by each method.
The proposed method could express both bright and dark areas for all exposure values.
In contrast, Yang et al. could not preserve the relative luminance, especially when the ratio of the exposure value is high like for $\pm 3$ EV or $\pm 4$ EV.
The proposed method could preserve the relative luminance even in such conditions.

\begin{figure}[t]
	\centering
	\includegraphics[width= 0.7\columnwidth]{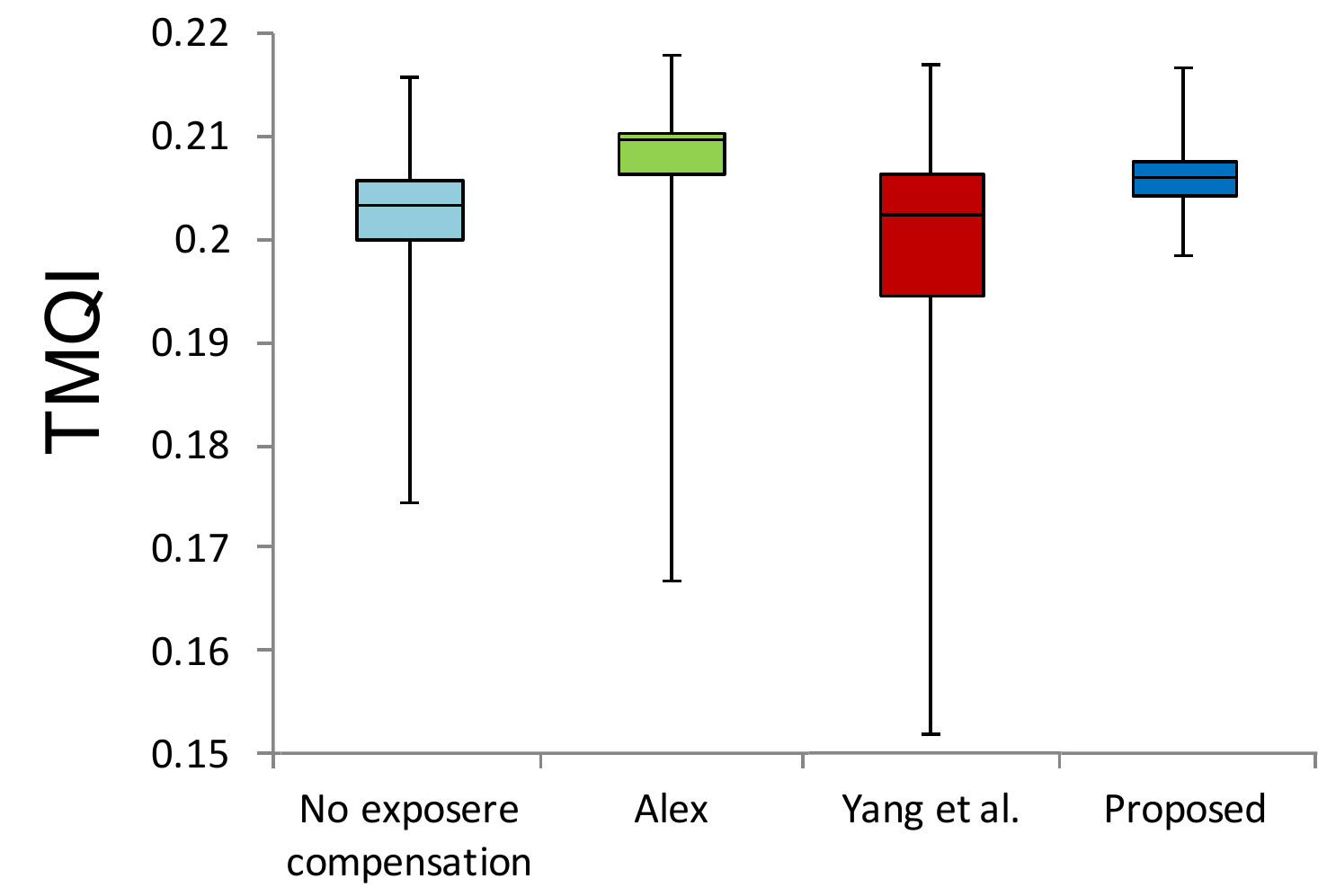}
	\caption{Experimental results (TMQI). Boxes span from the first to the third quartile referred to as $Q_1$ and $Q_3$, and whiskers show maximum and minimum values in the range of $[Q_1-1.5(Q_3-Q_1),Q_3+1.5(Q_3-Q_1)]$. Band inside box indicates median.}
	\label{fig:box_tmqi}
\end{figure}

\begin{figure}[t]
	\centering
	\includegraphics[width= 0.7\columnwidth]{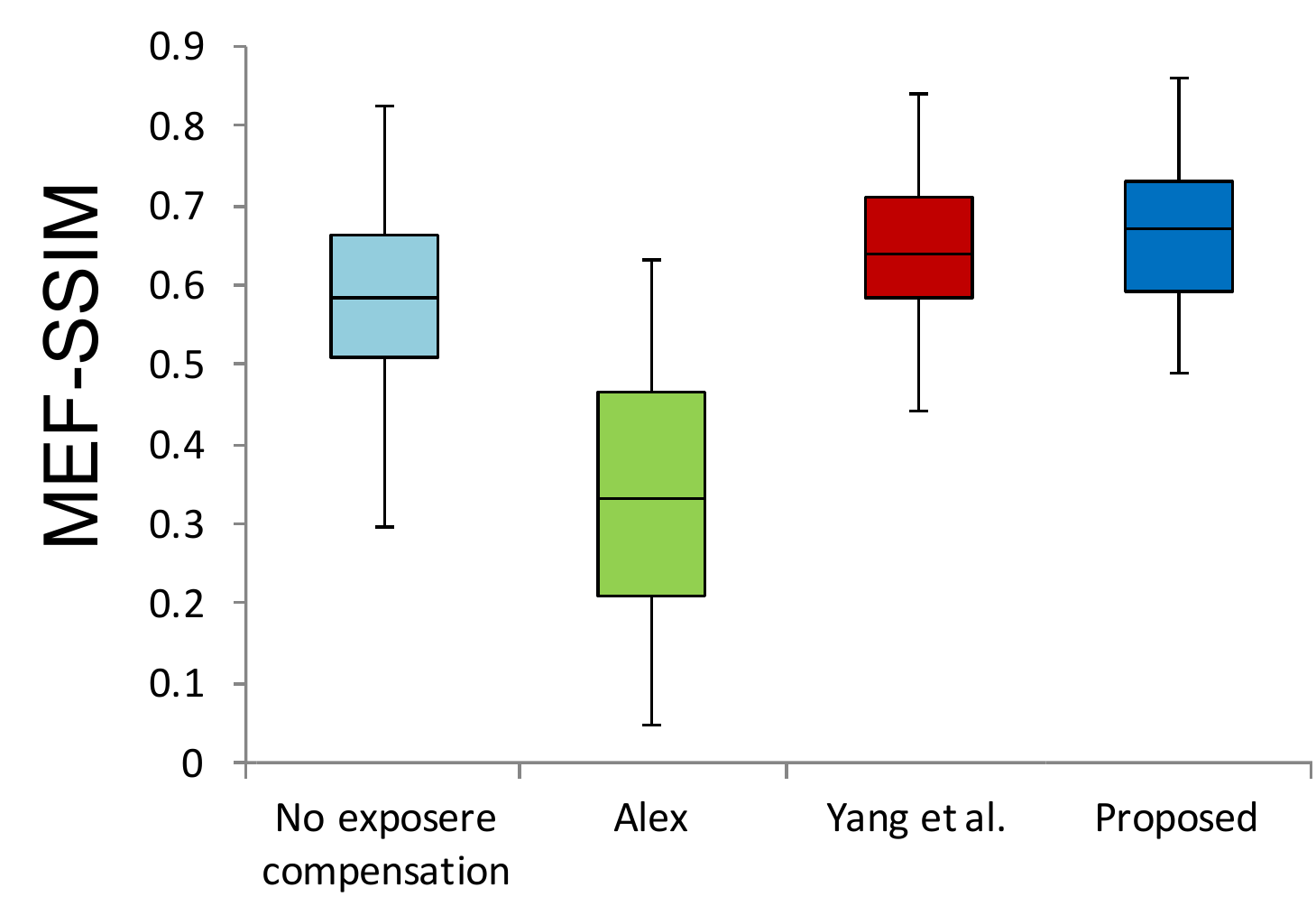}
	\caption{Experimental results (MEF-SSIM). Boxes span from the first to the third quartile referred to as $Q_1$ and $Q_3$, and whiskers show maximum and minimum values in the range of $[Q_1-1.5(Q_3-Q_1),Q_3+1.5(Q_3-Q_1)]$. Band inside box indicates median.}
	\label{fig:box_ssim}
\end{figure}

Figures \ref{fig:box_tmqi} and \ref{fig:box_ssim} show the box-plot of TMQI and MEF-SSIM scores for $28\times 4 = 112$ fused images.
From the results, the proposed method is demonstrated not only to have high scores but also to have narrow range.
The results mean the proposed method almost always provided good results under various conditions.
\begin{table}[t]
  \center
  \caption{Average scores of TMQI}
  \begin{tabular}{| l | c | r | r | r |} \hline
    				& $\pm 1{\rm EV}$ & $\pm 2{\rm EV}$ & $\pm 3{\rm EV}$ & $\pm 4{\rm EV}$ \\ \hline \hline
    No correction
    & 0.2061 & 0.2047 & 0.2012 & 0.1981 \\
    Alex \cite{alex}
    & 0.2051 & 0.2048 & 0.2033 & 0.2005 \\
    Yang et al. \cite{yang2018multi}
    & 0.2069 & 0.2034 & 0.1960 & 0.1882 \\
    Proposed&&&&\\(without local enhancement)
    & 0.2072 & 0.2065 & 0.2054 & 0.2044 \\ 
    Proposed&&&&\\(with local enhancement)
    & \bf0.2073 & \bf0.2066 & \bf0.2055 & \bf0.2045 \\
    Proposed&&&&\\(with Eqs. \eqref{eq:x2lum_R} and \eqref{eq:x2lum_B})
    & 0.2073 & 0.2065 & 0.2054 & 0.2043 \\  \hline
  \end{tabular}
  \label{tb:tmqi_all}
\end{table}

\begin{table}[t]
  \center
  \caption{Average scores of MEF-SSIM}
  \begin{tabular}{| l | c | r | r | r |} \hline
    				& $\pm 1{\rm EV}$ & $\pm 2{\rm EV}$ & $\pm 3{\rm EV}$ & $\pm 4{\rm EV}$ \\ \hline \hline
    No correction
    & 0.6346 & 0.6134 & 0.5698 & 0.5099 \\
    Alex \cite{alex}	
    & 0.3251 & 0.3250 & 0.3250 & 0.3248 \\
    Yang et al. \cite{yang2018multi}	
    & \bf0.6805 & \bf0.6772 & 0.6321 & 0.5848 \\
    Proposed&&&&\\(without local enhancement)
    & 0.6602 & 0.6293 & 0.6041 & 0.5924 \\ 
    Proposed&&&&\\(with local enhancement)	
    & 0.6666 & 0.6633 & \bf0.6564 & \bf0.6658 \\
    Proposed&&&&\\(with Eqs. \eqref{eq:x2lum_R} and \eqref{eq:x2lum_B})
    & 0.6652 & 0.6542 & 0.6477 & 0.6552 \\  \hline
  \end{tabular}
  \label{tb:ssim_all}
\end{table}

\subsection{Simulation with photographing}
\subsubsection{Dataset}
Photographs taken by Canon EOS 5D Mark $\rm I\hspace{-.1em}I$ camera were directly used as input images $\bf X$.
We also used Magic Lantern \cite{ml}, which is a firmware to perform dual-ISO sensing.
The shutter speed and the aperture were set by auto exposure of the camera at ISO 800.
This condition means that the exposure value is 0 EV at ISO 800. For the dual-ISO imaging, ISO 200 and ISO 3200 correspond to -2 EV and +2 EV respectively.
We used nine dual-ISO sensing images.

\subsubsection{Objective metrics}
In this experiment, there are no ideal images to be reference images.
Thus, we used discrete entropy and statistical naturalness as objective quality metrics, which do not require any reference images.
Discrete entropy represents the amount of information in an image.
For each score, a larger value means higher quality.

\subsubsection{Result}
\begin{figure*}[t]
	\centering
	\subfigure[No exposure compensation ($\pm 1\rm EV$)]{
		\label{subfig:dual-b}
		\includegraphics[width = 0.5\columnwidth]{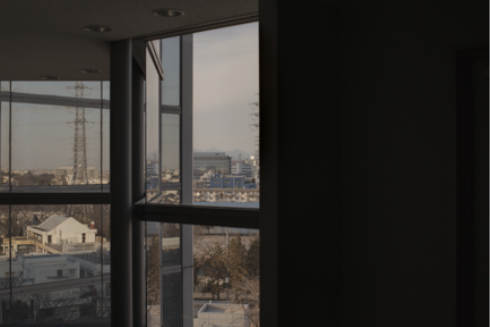}}
	\subfigure[Alex ($\pm 1\rm EV$)]{
		\label{subfig:dual-b}
		\includegraphics[width = 0.5\columnwidth]{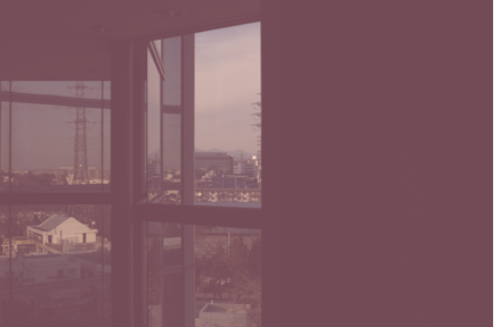}}
	\subfigure[Yang et al. ($\pm 1\rm EV$)]{
		\label{subfig:dual-c}
		\includegraphics[width = 0.5\columnwidth]{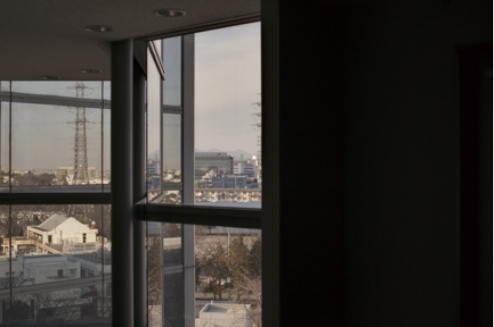}}
	\subfigure[Proposed method ($\pm 1\rm EV$)]{
		\label{subfig:dual-a}
		\includegraphics[width = 0.5\columnwidth]{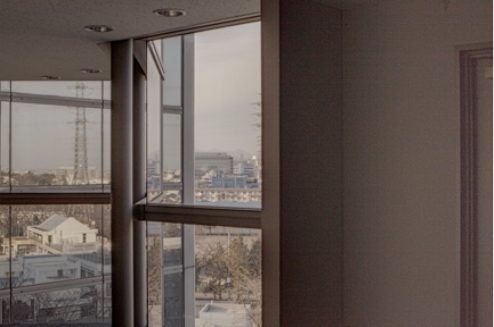}}
	\hspace{20mm}
	\subfigure[No exposure compensation ($\pm 2\rm EV$)]{
		\label{subfig:dual-e}
		\includegraphics[width = 0.5\columnwidth]{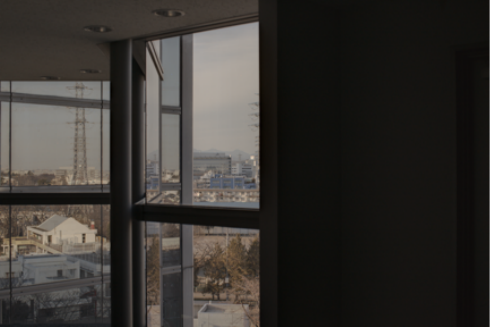}}
	\subfigure[Alex ($\pm 2\rm EV$)]{
		\label{subfig:dual-e}
		\includegraphics[width = 0.5\columnwidth]{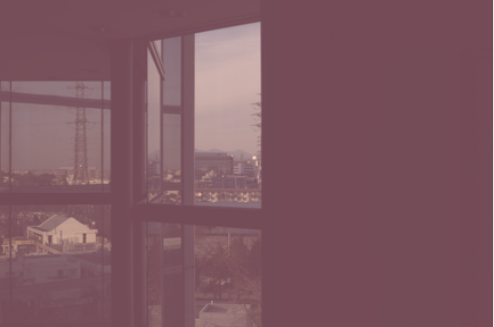}}
	\subfigure[Yang et al. ($\pm 2\rm EV$)]{
		\label{subfig:dual-f}
		\includegraphics[width = 0.5\columnwidth]{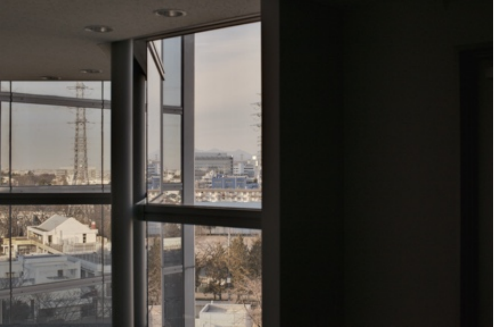}}
	\subfigure[Proposed method ($\pm 2\rm EV$)]{
		\label{subfig:dual-d}
		\includegraphics[width = 0.5\columnwidth]{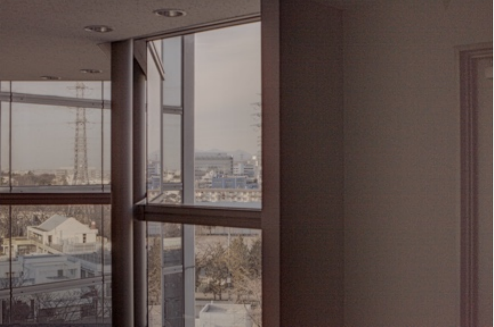}}
	\hspace{20mm}
	\subfigure[No exposure compensation ($\pm 3\rm EV$)]{
		\label{subfig:dual-h}
		\includegraphics[width = 0.5\columnwidth]{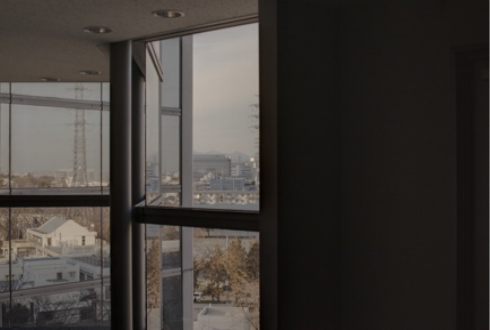}}
	\subfigure[Alex ($\pm 3\rm EV$)]{
		\label{subfig:dual-e}
		\includegraphics[width = 0.5\columnwidth]{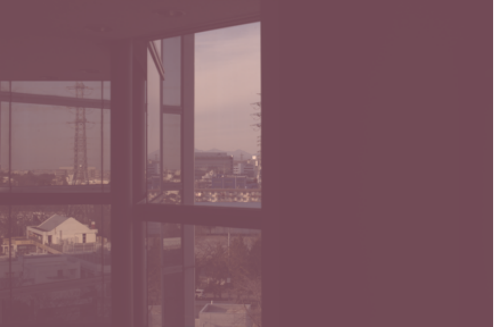}}
	\subfigure[Yang et al. ($\pm 3\rm EV$)]{
		\label{subfig:dual-i}
		\includegraphics[width = 0.5\columnwidth]{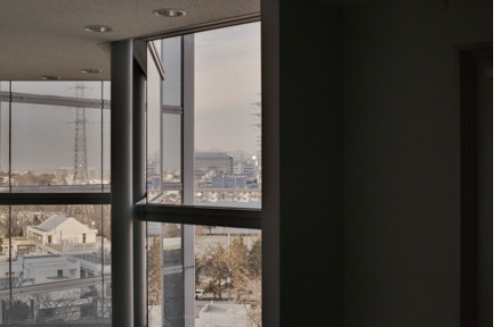}}
	\subfigure[Proposed method ($\pm 3\rm EV$)]{
		\label{subfig:dual-g}
		\includegraphics[width = 0.5\columnwidth]{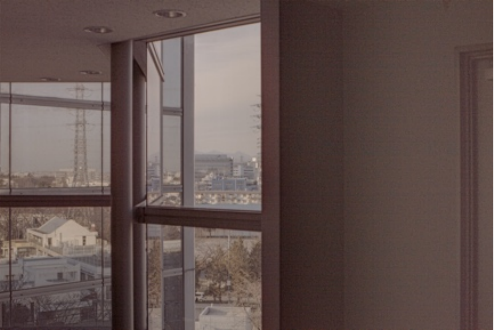}}
	\caption{Examples of fused images ${\bf Y}_{\rm out}$ $(S=5)$}
	\label{fig:result_2}
\end{figure*}

Tables \ref{tb:sn_all} and \ref{tb:entropy_all} denote the average scores of nine images.
From Table \ref{tb:sn_all}, the proposed method had a high score for all exposure values.
From Table \ref{tb:entropy_all}, in the case of $\pm1$ EV and $\pm2$ EV, the proposed method had higher scores than conventional methods, although Yang et al. had higher discrete entropy scores than the proposed method in the case of $\pm3$ EV.
Figure \ref{fig:result_2} shows examples of output images ${\rm Y}_{out}$ produced by each method.
Compared with the conventional methods, the proposed imaging successfully represent information of dark area for all exposure values.
In contrasts, Yang's method does not represent information of dark area.

\begin{figure}[t]
	\centering
	\includegraphics[width= 0.9\columnwidth]{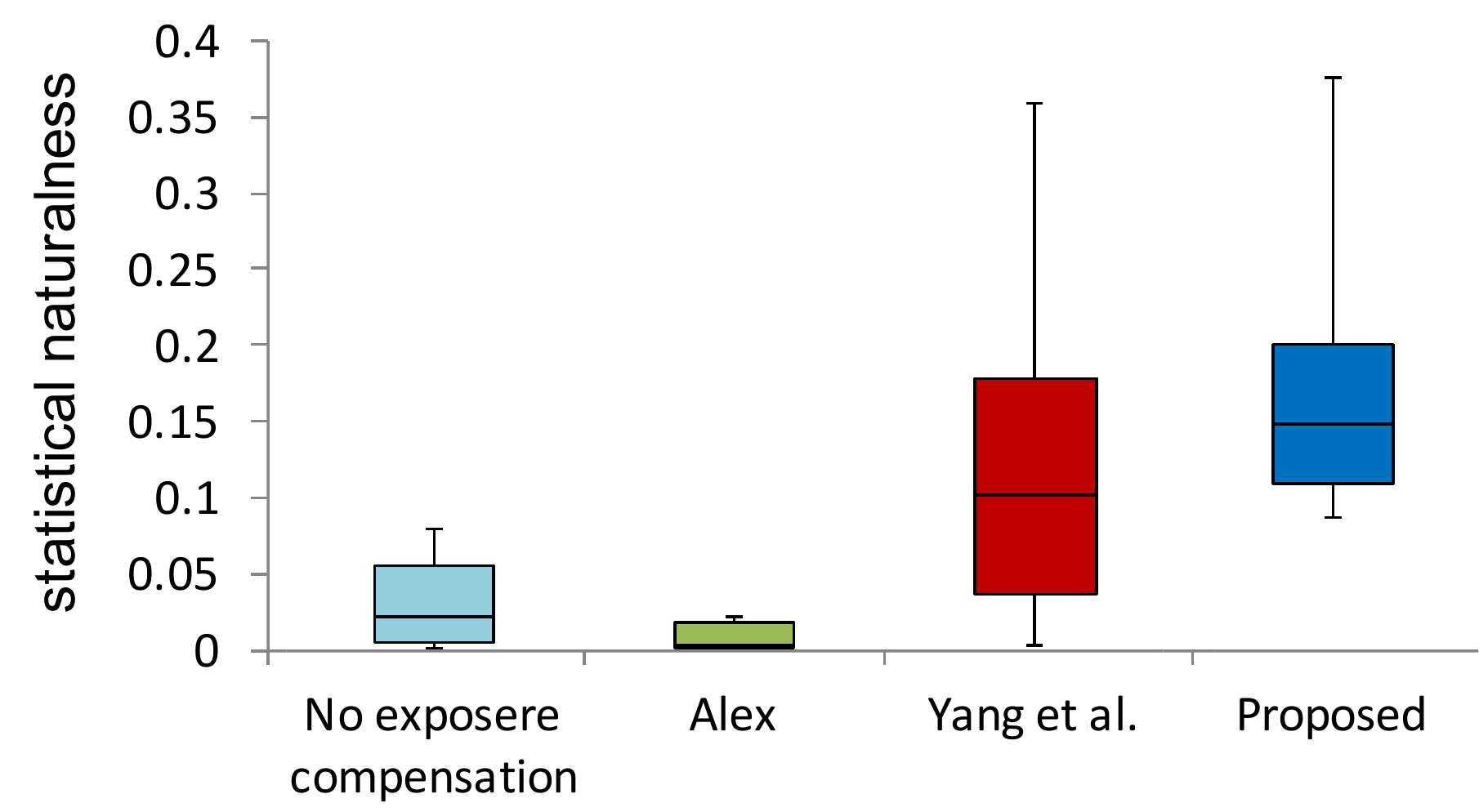}
	\caption{Experimental results (statistical naturalness). Boxes span from the first to the third quartile referred to as $Q_1$ and $Q_3$, and whiskers show maximum and minimum values in the range of $[Q_1-1.5(Q_3-Q_1),Q_3+1.5(Q_3-Q_1)]$. Band inside box indicates median.}
	\label{fig:box_sn}
\end{figure}

\begin{figure}[t]
	\centering
	\includegraphics[width= 0.8\columnwidth]{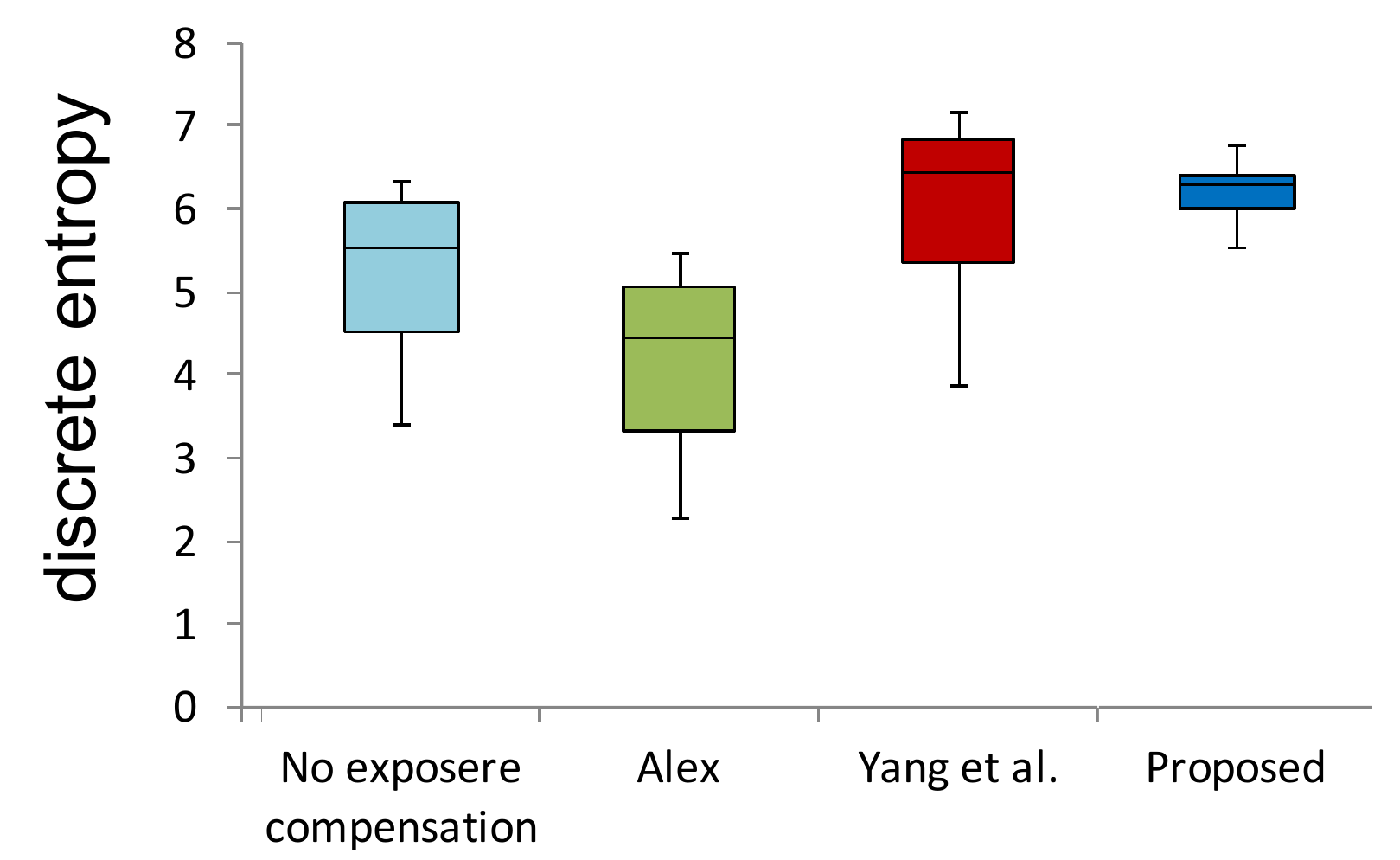}
	\caption{Experimental results (discrete entropy). Boxes span from the first to the third quartile referred to as $Q_1$ and $Q_3$, and whiskers show maximum and minimum values in the range of $[Q_1-1.5(Q_3-Q_1),Q_3+1.5(Q_3-Q_1)]$. Band inside box indicates median.}
	\label{fig:box_de}
\end{figure}

Figures \ref{fig:box_sn} and \ref{fig:box_de} show the box-plot of statistical naturalness and discrete entropy for $9 \times 3 = 27$ fused images.
The performance of the proposed method is demonstrated to offer high quality images under various conditions as well as in Figs.\ref{fig:box_tmqi} and \ref{fig:box_ssim}.

\begin{table}[t]
  \center
  \caption{Average scores of statistical naturalness}
  \begin{tabular}{| l | c | r | r  |} \hline
    				& $\pm 1{\rm EV}$ & $\pm 2{\rm EV}$ & $\pm 3{\rm EV}$  \\ \hline \hline
    No correction	& 0.0229 & 0.0303 & 0.0385 \\
    Alex \cite{alex}	& 0.0082 & 0.0078 & 0.0075 \\
    Yang et al. \cite{yang2018multi}	& 0.0794 & 0.1336 & 0.1466 \\
    Proposed&&&\\
     (without local enhancement)& 0.1722 & 0.1469 & 0.1458 \\
    Proposed&&&\\
    (with local enhancement)& \bf0.1885 & \bf0.1579 & \bf0.1585 \\ \hline
  \end{tabular}
  \label{tb:sn_all}
\end{table}

\begin{table}[t]
  \center
  \caption{Average scores of discrete entropy}
  \begin{tabular}{| l | c | r | r  |} \hline
    				& $\pm 1{\rm EV}$ & $\pm 2{\rm EV}$ & $\pm 3{\rm EV}$  \\ \hline \hline
    No correction	& 5.0301 & 5.2525 & 5.5203 \\
    Alex \cite{alex}	& 4.1619 & 4.1092 & 4.1349 \\
    Yang et al. \cite{yang2018multi}	& 5.5740 & 6.1244 & \bf6.5076 \\
     Proposed&&&\\
     (without local enhancement)& 6.3579 & 6.1785 & 6.0886\\
    Proposed&&&\\
    (with local enhancement)& \bf6.3610 & \bf6.1886 & 6.0997 \\ \hline
  \end{tabular}
  \label{tb:entropy_all}
\end{table}


\section{Conclusion}
In this paper, we proposed a new image fusion method for single-shot imaging with SVE.
In the single-shot imaging with SVE, the number of exposures is generally limited to two.
To improve the limitation, the exposure compensation for input images is automatically performed so that generated multi-exposure images clearly show all regions in a scene.
In the proposed method, generated multi-exposure images can be applied to any MEF methods.
We evaluated the effectiveness of the proposed method in terms of four objective quality metrics: TMQI, MEF-SSIM, statistical naturalness, and discrete entropy.
Experimental results showed that the proposed method can produce high-quality images compared with conventional fusion methods.
\bibliographystyle{ieicetr}

\profile{Chihiro Go}{received his B.Eng. degree from Tokyo University of Agriculture and Technology, Japan, in 2017. He is a Master course student at Tokyo Metropolitan University, Japan. His research interests are in the area of image processing.}
\label{profile}

\profile{Yuma Kinoshita}{received his B.Eng. and M.Eng. degrees from Tokyo Metropolitan University, Japan, in 2016 and 2018, respectively. From 2018, he has been a Ph.D. student at Tokyo Metropolitan University. He received IEEE ISPACS Best Paper Award in 2016, IEEE Signal Processing Society Japan Student Conference Paper Award in 2018, and IEEE Signal Processing Society Tokyo Joint Chapter Student Award in 2018, respectively. His research interests are in the area of image processing. He is a student member of IEEE and IEICE.}
\label{profile}

\profile{Sayaka Shiota}{received B.E., M.E., and Ph.D. degrees in intelligence and computer science, engineering, and engineering simulation from the Nagoya Institute of Technology, Nagoya,
Japan in 2007, 2009, and 2012, respectively. From February 2013 to March 2014, she worked as a Project Assistant Professor at the Institute of Statistics Mathematics. In April of 2014, she joined Tokyo Metropolitan University as an Assistant Professor. Her research interests include statistical speech recognition and speaker verification. She is a member of the Acoustical Society of Japan (ASJ), the IEICE, the ISCA, APSIPA, and the IEEE.}
\label{profile}

\profile{Hitoshi Kiya}{received his B.E and M.E. degrees from Nagaoka University of Technology, in 1980 and 1982 respectively, and his Dr. Eng. degree from Tokyo Metropolitan University in 1987. In 1982, he joined Tokyo Metropolitan University, where he became Full Professor in 2000. From 1995 to 1996, he attended the University of Sydney, Australia as a Visiting Fellow. He is a Fellow of IEEE, IEICE and ITE. He currently serves as President of APSIPA, and he served as Inaugural Vice President (Technical Activities) of APSIPA in 2009-2013, and as
Regional Director-at-Large for Region 10 of IEEE Signal Processing Society in 2016-2017. He was also President of IEICE Engineering Sciences Society in 2011-2012, and he served there as Vice President and Editor-in-Chief for IEICE Society Magazine and Society Publications. He was Editorial Board Member of eight journals, including IEEE Trans. on Signal Processing, Image Processing, and Information Forensics and Security, Chair of two technical committees and Member of nine technical committees including APSIPA Image, Video, and Multimedia Technical Committee (TC), and IEEE Information Forensics and Security TC. He has organized a lot of international conferences, in such roles as TPC Chair of IEEE ICASSP 2012 and as General Co-Chair of IEEE ISCAS 2019. Dr. Kiya is a recipient of numerous awards, including six best paper awards.}

\end{document}